\documentclass[useAMS,usenatbib,letters]{mn2e}

\usepackage{graphicx}
\usepackage{txfonts}

\usepackage{natbib}

\newcommand{\mnras}[1]{MNRAS}
\newcommand{\apj}[1]{ApJ}
\newcommand{\apjl}[1]{ApJL}
\newcommand{\apjs}[1]{ApJS}
\newcommand{\aj}[1]{AJ}
\newcommand{\aap}[1]{A\& A}

\newcommand{\mue}{$\mu_e$}

\title[The Fudamental Plane as a Confounding Correlation]{The Fundamental Plane of Early-Type Galaxies as a Confounding Correlation}
 \author[D. Fraix-Burnet]{D. Fraix-Burnet$^{1}$\thanks{E-mail:
 fraix@obs.ujf-grenoble.fr}\\
$^{1}$Universit\'e Joseph Fourier - Grenoble 1 / CNRS, Institut de Plan\'etologie et d'Astrophysique de Grenoble, BP 53, F-38041 Grenoble cedex 9, France}

\begin{document}

\date{Accepted 2011 June 12.  Received 2011 May 17; in original form 2011 March 25.}

\pagerange{\pageref{firstpage}--\pageref{lastpage}} \pubyear{2011}

\maketitle

\label{firstpage}

\begin{abstract}
Early-type galaxies are characterized by many scaling relations. One of them, the so-called fundamental plane is a relatively tight correlation between three variables, and has resisted a clear physical understanding despite many years of intensive research. Here, we show that the correlation between the three variables of the fundamental plane can be the artifact of the effect of another parameter influencing all, so that the fundamental plane may be understood as a confounding correlation. Indeed, the complexity of the physics of galaxies and of their evolution suggests that the main confounding parameter must be related to the level of diversification reached by the galaxies. Consequently, many scaling relations for galaxies are probably evolutionary correlations.
\end{abstract}


\section{Introduction}
\label{introduction}

Galaxies are huge systems of stars, gas and dust, that have assembled through a complex history of collapses, internal perturbations, interactions, mergers and even ejections of some material. This complexity renders quite remarkable the scaling relations found between many observables and properties, like size, mass, surface brightness, velocity dispersion, magnitude, color, metallicity, black hole mass or spectral features. The fundamental plane is a famous correlation between three variables, the effective radius, the central velocity dispersion and the surface brightness within the effective radius \citep{Dressler1987,Djorgovski1987,Nieto1990}. For more than 20 years of investigations, it has resisted a satisfactory physical understanding \citep{Robertson2006, DOnofrio2008, Gargiulo2009, Graves2009, Nigoche-Netro2009, Fraix2010}.

For (astro)physicists, such scaling relations reveal some underlying physical laws that are described by a set of equations. However, for statisticians, correlations are not always causal, as lurking or confounding parameters can do the trick. In general, these correlations are called "spurious" although this more specifically concerns acknowledged false correlations.

A famous example of a "spurious" correlation in statistics is the relation between ice cream sales and number of drownings. The confounding parameter is the high temperature which favours both the ice cream consumption and the number of people swimming, hence the probability of drownings. So even unrelated phenomena can appear correlated. Other examples, like the relation between the size of a children and his school level or between the distance of the Pioneer probe to the Earth and the average Earth temperature since its launch, have time or evolution as confounding factor.

Indeed the correlation between children's size and school level is statistical: the taller ones being more probably found at higher levels. This is reminiscent of a classical reasoning used in the determination of the cosmic distance ladder: for example, more luminous stars are more probably found in bigger galaxies. Simillarly, because of evolution, bigger galaxies are  
more probably more massive, more luminous, more metallic, with a higher velocity dispersion, etc. 
Hence statistically, a lot of properties must appear correlated, the confounding parameter being the level of diversification.

In this paper, we show that correlations between the three variables of the fundamental plane can easily be the artifact of the effect of another parameter influencing all. In other words, 
we show that the fundamental plane may be understood as a confounding correlation. We propose that the confounding parameter(s) is (are) related to evolution. Consequently, many scaling relations of galaxies are probably evolutionary correlations.

In Sect.~\ref{spurcorr} we derive the conditions for parametric relations to yield a planar correlation in a 3-variable space. We then dedicate a full section to the fundamental plane (Sect.~\ref{FP}) to show how easily these conditions can be fulfilled. We finally discuss why and how the evolution of galaxies is probably the confounding parameter (Sect.~\ref{evolutionary}).

\section[]{Confounding correlations}
\label{spurcorr}

Let us consider three variables $\Omega_1, \Omega_2, \Omega_3$ that are all functions of a same generic parameter $\widetilde{X}$. For sake of simplicity in this paper, let us assume that these functions are power laws:

\begin{equation}
\left\{
\begin{array}{rcl}
    \Omega_1 & = & A_1\widetilde{X}^p \\
    \Omega_2 & = & A_2\widetilde{X}^s \\
    \Omega_3 & = & A_3\widetilde{X}^t 
\end{array}
\right.
\label{eq:powerlawsgen}
\end{equation} 

with $A_1$, $A_2$ and $A_3$ being constant. Any linear correlation of the form
\begin{equation}
   \log\Omega_1 = a \log\Omega_2 + b \log\Omega_3 + c
\label{eq:fundplanegen}
\end{equation} 

where $a$, $b$ and $c$ are constant, is thus expressed as

\begin{equation}
   p \log \widetilde{X} + \log A_1= as\log \widetilde{X} + b t \log \widetilde{X}  + a\log A_2 + b\log A_3 + c .
\label{eq:fprewritesgen}
\end{equation}

This expression should be valid for all $\widetilde{X}$, implying that

\begin{equation}
\left\{
\begin{array}{rcl}
p &=& a s  +  b t    \\
 \log A_1 &=& a\log A_2 + b\log A_3 + c.   
\end{array}
\right.
\label{eq:condition1gen}
 \end{equation}

This set of two linear equations generally yields solutions for $a$ and $b$. The result is thus a plane in the 3-D space $\left( \Omega_1, \Omega_2, \Omega_3 \right)$.

The generic parameter $\widetilde{X}$ can be a multivariate component, making the parametric dependence of $\Omega_1, \Omega_2$ and $\Omega_3$ mutlivariate. For instance, consider power laws with two parameters, $X_1$ and $X_2$:

\begin{equation}
\left\{
\begin{array}{rcl}
    \Omega_1 & = & A_1 X_1^p X_2^{p^\prime} \\
    \Omega_2 & = & A_2 X_1^s X_2^{s^\prime} \\
    \Omega_3& = & A_3 X_1^t X_2^{t^\prime} 
\end{array}
\right.
\label{eq:powerlawsmult}
\end{equation} 

Then it is easy to show that a correlation like equation~(\ref{eq:fundplanegen}) holds if:
\begin{equation}
\left\{
\begin{array}{rcl}
p &=& a s +  b t  \\
p^\prime &=& a s^\prime  + b  t^\prime   \\
 \log A_1 &=& a\log A_2 + b\log A_3 + c.  
\end{array}
\right.
\label{eq:condition1mult} 
 \end{equation} 

This set of equations is more constraining than equation~(\ref{eq:condition1gen}). But if for instance the second relation in equation~(\ref{eq:condition1mult}) is not exactly fulfilled, then the parameter $X_2$ can be considered as noise adding a dispersion to the correlation defined by the two other relations. In other words, $X_2$ could generate a thickness to the plane defined by equation~(\ref{eq:fundplanegen}).

To be complete, we must discuss the physics in the parameter $\widetilde{X}$. This parameter is supposed to influence all the three variables. We distinguish two possbilities:
causality or evolution. In the first case, the relation between the variables and $\widetilde{X}$ is driven by direct causality, that is some parameter influences directly each of the three variables $\Omega_1, \Omega_2, \Omega_3$ through physical laws. This is illustrated in Sect.~\ref{virial}. In the second case, the relation is statistical in the sense that each variable is bound to evolve so that even totally unrelated variables can show an apparent correlation. This is discussed in Sect.~\ref{novirial} and Sect.~\ref{evolutionary}.

\section[]{The fundamental plane of early-type galaxies}
\label{FP}

\subsection{General constraint}
\label{general}

Let us put $\Omega_1 = r_e$ the effective radius, $\Omega_2 = \sigma$ the velocity dispersion, and $\Omega_3 = L$ the luminosity. equation~(\ref{eq:powerlawsgen}) becomes:

\begin{equation}
\left\{
\begin{array}{rcl}
     r_e    & = & A_1  \widetilde{X}^p \\
  \sigma & = & A_2  \widetilde{X}^s \\
           L &= & A_3   \widetilde{X}^t 
\end{array}
\right.
\label{eq:powerlaws}
\end{equation} 

The surface brightness \mue\ can be expressed as 
\begin{equation} 
\begin{array}{rcl}
\mu_e &=&  -2.5 \log\left(L/4 \pi r_e^2\right) + m  \\
         &=& \left( -2.5 t + 5 p \right) \log \widetilde{X} + 2.5 \log (4 \pi) + m
\end{array}
\label{eq:brie}
\end{equation} 

where $m$ is a constant of normalisation. Any linear correlation of the form

\begin{equation}
   \log r_e = a \log\sigma + b \mu_e + c
\label{eq:fundplane}
\end{equation} 

 translates to 
\begin{eqnarray}
   p \log \widetilde{X} + \log A_1= as\log \widetilde{X} + b \left( -2.5 t + 5 p \right) \log \widetilde{X}  + \nonumber \\
a\log A_2 + b\left( 2.5\log\left(4\pi A_1^2/A_3\right) + m \right) + c .
\label{eq:fprewrites}
\end{eqnarray} 

This implies

\begin{equation}
\left\{
\begin{array}{rcl}
p &=& s a + \left(-2.5 t + 5 p\right)  b    \\
log A_1  &=& a\log A_2 + b\left( 2.5\log\left(4\pi A_1^2/A_3\right) + m \right) + c. 
\end{array}
\right.
\label{eq:condition1}
\end{equation}

If a solution can be found for $a$ and $b$ from equation~(\ref{eq:condition1}), then the equation of the fundamental plane equation~(\ref{eq:fundplane}) is obtained. There is no need of any further assumption to explain the fundamental plane. This demonstration is made here with a simple power-law assumption in equation~(\ref{eq:powerlaws}). But this result is also true for more complex functions, with equation~(\ref{eq:condition1}) being replaced by more complicated conditions. 

To understand the origin equation~(\ref{eq:fundplane}), we need to solve equation~(\ref{eq:condition1}). Since there are too many unknowns, additional input is required. Two approaches are possible: either input some a priori knowledge to determine the functions of $\widetilde{X}$ and derive coefficients $a$ and $b$ (Sect.\ref{virial}), or conversely use the observations to determine $a$ and $b$ and derive constraints on the functions of $\widetilde{X}$ (Sect.\ref{novirial}).

\subsection{A priori physical input}
\label{virial}

In this section, we consider the input of a priori knowledge believed to be relevant for the physics of galaxies. At first glance, it is quite logical to consider that mass, either dynamical, true or stellar, is somehow related to the radius, the velocity dispersion and the surface brightness, essentially because it influences the density of stars and their kinematics. So we consider $\widetilde{X}=M$ in equation~(\ref{eq:powerlaws}).

One obvious way to link the kinematics to the mass is through the virial theorem. Hence, let us assume that the virial equation holds and that the ratio between the average squared velocity and $\sigma^2$, and the ratio between $r_e$ and the gravitational radius, are constant. This gives
\begin{equation}
   \sigma^2 r_e \propto M \Rightarrow 2s+p=1
\label{eq:psvirial}
\end{equation}

Using equation~(\ref{eq:condition1}) and equation~(\ref{eq:psvirial}) we obtain
\begin{eqnarray}
&&     1-2s  = sa + \left(-2.5 t + 5 \left( 1-2s\right) \right)  b   \nonumber\\
\Rightarrow &&s \left( a+2-10b\right) =2.5 t b - 5 b +1 \nonumber\\
\Rightarrow &&\left\lbrace 
\begin{array}{l}
2.5 t b - 5 b +1=0 \hfill \mbox{ if } a+2-10b = 0 \\
s= \left( 2.5 t b - 5 b +1\right) /\left(a+2-10b \right)  \>\> \mbox{ otherwise. }
\end{array}
\right. 
\label{eq:psvirial2}
\end{eqnarray} 

Unfortunately, the brightness has no direct relation to mass, but it might be assumed that the mass is essentially due to the stars that are responsible for the luminosity. If we also assume that the ratio $M/L$ is constant for a given population of galaxies, we must have:
\begin{equation}
   L \propto M \Rightarrow t=1.
\label{eq:MsLcst}
\end{equation} 

Replacing this value of $t$ in equation~(\ref{eq:psvirial2}), one obtains two solutions:

\begin{eqnarray}
&(i)&\left\lbrace 
\begin{array}{l}
2.5  b - 5 b +1=0 \mbox{ and } a+2-10b = 0 \\
\;\;\; \Rightarrow b=0.4 \mbox{ and } a=2 \\
\end{array}
\right. \nonumber\\
\mbox{ or } \\
&(ii)&\left\lbrace 
\begin{array}{l}
s= \left( 2.5 b - 5 b +1\right) /\left(a+2-10b \right) \\
\;\;\; \Rightarrow s= \left( 1-2.5 b\right) /\left(a+2-10b \right) 
\end{array}
\right. \nonumber
\label{eq:psvirial3}
\end{eqnarray}

The first solution is well known, being the so-called virial plane $a=2$ and $b=0.4$, a classical solution obtained with the assumptions made to derive equation~(\ref{eq:psvirial}) and equation~(\ref{eq:MsLcst}). In this case the dependency of $\sigma$ (or $r_e$) on the mass $M$ is undetermined ($s$ and $p$ cannot be computed). 

However, the second solution yields different values of the coefficients $a$ and $b$ with the same classical assumptions. In this case, the dependency of $r_e$ or $\sigma$ on $M$ can be computed. Hence the tilt of the observed fundamental plane with respect to the virial plane could simply be explained in this way whithout relaxing any of the assumptions made above. Taking typical observed values $a\simeq 1.1$ and $b\simeq 0.3$ \citep[e.g.][]{Fraix2010}, we find $s\simeq 2.5$ and $p\simeq -4$. This yields strong dependences of the variables on mass. In particular, the $r_e$ vs $M$ relation is steeper than that found for instance in numerical simulations by \citet{Robertson2006}.

As a conclusion, the two possible solutions implied by equation~(\ref{eq:psvirial3}), leading to $s$ and $p$ either undefined or relatively high, can be debated, but it appears clearly that the assumptions made to derive the virial plane 
are a very restrictive case of more general conditions (equation~(\ref{eq:condition1})) that lead to a fundamental plane correlation. Interestingly, one usual modification to explain the apparent tilt of the observed fundamental  plane with respect to the virial plane is to relax the assumption in equation~(\ref{eq:MsLcst}) with $t\ne1$ and rely on the observations for this value. In the next section we show that we can go a step further and use the more general conditions in equation~(\ref{eq:condition1}) with the sole observations as constraints without any a priori input from the virial hypothesis. This allows the exploration of wider physics: the virial conditions (equations~(\ref{eq:psvirial}) and (\ref{eq:MsLcst})), if fulfilled, should be derived from the observations, not taken as an a priori.

\subsection{Observational constraints}
\label{novirial}

Henceforth, let us use the relations in equation~(\ref{eq:condition1}) to derive the constraints on the functions in equation~(\ref{eq:powerlaws}) from the observations. To derive the three exponents $s$, $p$ and $t$, we take the values for $a$ and $b$ as provided by the observed fundamental plane correlations, and also other observed correlations. 

As mentioned above, the observed fundamental plane yields typical values of $a\simeq 1.1$, $b\simeq 0.3$ and $c=-8.5$ (and $m=4.45$ in the R band). However, these are known to depend on the sample, the wavelength, the redshift and more generally on the group of galaxies \citep[e.g.][]{Fraix2010}. This strongly indicates that the functions of $\widetilde{X}$ in equation~(\ref{eq:powerlaws}) and $\widetilde{X}$ itself are not universal and depend on the population of galaxies and their assembly histories. This can be seen in the numerical simulations by \citet{Robertson2006} where the $r_e$ vs $M$ relations change according to the sequences and nature of the merging events. 

With the observed values for $a$, $b$ and $c$, equation~(\ref{eq:condition1}) becomes

\begin{eqnarray}
   &&
\left\{
\begin{array}{rcl}
p &\simeq& 1.1s  + \left(-2.5 t + 5 p\right)  0.3    \\
\log A_1  &\simeq& 1.1\log A_2 + 0.3\left( 2.5\log\left({4\pi A_1^2}/{A_3}\right) + 4.45 \right) \\ && \hfill - 8.5
\end{array}
\right.
\nonumber\\ \Rightarrow
&&\left\{
\begin{array}{rcl}
p &\simeq& - 2.2s + 1.5t     \\
A_1 &\simeq& 2\ 10^{-13}\ A_2^{-2.2}\ A_3^{1.5}.
\end{array}
\right.
\label{eq:condition1obs}
\end{eqnarray}

The above set of relations is independent of the actual meaning of $\widetilde{X}$. The physics can here come to our help to consider different possibilities for the confounding parameter $\widetilde{X}$, but here we still only use observations (of real objects or synthetic ones from numerical simulations) to determine the shape of the functions of $\widetilde{X}$.

Let us consider the fig.~5 in \citet{Hopkins2008} that gives the dependence of $r_e$ and $\sigma$ on the mass fraction $f_{starburst}$ of the starburst which is an indicator of the level of dissipation in passed merger. The dissipation has been found to be a key element to explain the very existence of the fundamental plane. 
 In simple words, without dissipation, the merger product would not follow a plane, that is to say that the correlation of equation~(\ref{eq:fundplane}) would not exist. Hence if we assume that the relations given in their fig.~5 are observations (of synthetic populations of galaxies in this case), then we derive that $p\simeq -1$ and $s\simeq 1$. We then compute the exponent $t$ from equation~(\ref{eq:condition1obs}) and find $t \simeq 2.1$. We thus have

\begin{equation}
\left\{
\begin{array}{rcl}
     r_e    & \propto & f_{starburst}^{-1} \\
  \sigma & \propto & f_{starburst}^{1} \\
           L &\propto &  f_{starburst}^{0.8} \\
\end{array}
\right.
\label{eq:relfstar}
\end{equation} 

These functions are in agreement with the expected role of dissipation: evacuates angular momentum, concentrates the matter in the bulge ($r_e$ diminishes), increases the surface brightness, and increases the velocity dispersion. 

Let us consider another candidate for $\widetilde{X}$: $M_{BH}$ the mass of the central black hole. It can be guessed that it could play a key role in the evolution of the properties of galaxies in the course of diversification. On the Fig.~4 in \citet{Hopkins2009}, there is a clear dependence of $r_e$ and $\sigma$ on $M_{BH}$. The relations are not exactly linear, but let us assume that it is the case to a first approximation. We find that $p\simeq 0.63$ and $s\simeq 0.28$, implying from equation~(\ref{eq:condition1obs}) that $t\simeq 0.83$. Then

\begin{equation}
\left\{
\begin{array}{rcl}
     r_e    & \propto & M_{BH}^{0.63} \\
  \sigma & \propto & M_{BH}^{0.28} \\
           L &\propto &  M_{BH}^{0.83} \\
\end{array}
\right.
\label{eq:relMBH}
\end{equation}

These values compare very favorably to the results obtained by \citet{Gultekin2009} ($\sigma \propto M_{BH}^{0.25}$ and  $L \propto  M_{BH}^{0.9}$) and \citet{Marconi2003} ($L \propto  M_{BH}^{0.83}$) for elliptical galaxies. We then conclude that the fundamental plane is explained by solely assuming that the black hole mass evolves in parallel with other properties of galaxies, without any direct physical modeling of each relation in equation~(\ref{eq:relMBH}),

Finally, it is interesting to remark that we can combine the two examples above with $\widetilde{X}=\left(f_{starburst},M_{BH}\right)$ like in equation~(\ref{eq:powerlawsmult}) and obtain the conditions equation~(\ref{eq:condition1mult}). 
We then obtain exactly a combination of equation~(\ref{eq:relfstar}) and equation~(\ref{eq:relMBH}):

\begin{equation}
\left\{
\begin{array}{rcl}
     r_e    & \propto & f_{starburst}^{-1}\ M_{BH}^{0.63} \\
  \sigma & \propto & f_{starburst}^{1} \ M_{BH}^{0.28} \\
           L &\propto &  f_{starburst}^{2.1}\ M_{BH}^{0.83} \\
\end{array}
\right.
\label{eq:relcomb}
\end{equation} 

This solution is given for illustration only because one might question the independence between $f_{starburst}$ and $M_{BH}$ (see below).

We have thus proven that the fundamental plane correlation equation~(\ref{eq:fundplane}) can be easily and plausibly obtained with confounding parameters like merger dissipation or central black hole mass, without any assumption linked to the virial equilibrium.

\section[]{Evolutionary correlations}
\label{evolutionary}

In the course of diversification, many properties of galaxies change, and they tend to statistically change in a more or less monotonous way. For instance mass and metallicity are both bound to increase with the complexity of a galaxy assembly history, so that they appear to be statistically correlated. It seems difficult to avoid the evolution to act as a confounding factor.

We thus propose that the main confounding parameter is $\widetilde{X}=T$ with $T$ an indicator of the level of diversification, being something like an evolutionary clock not necessarily easily related to time or redshift.

In the simple case of a single stellar population, the time since formation is naturally a good evolutionary clock for some parameters like luminosity $L$, color and metallicity. However this is less obvious for $r_e$ or $\sigma$. Time since formation can probably be as good for a homogeneous population of galaxies if they are not affected by significant transformation events (interactions, mergers...). However, galaxies from the same homogeneous population (identical properties) do not form at the same epoch in the Universe, so that the time since formation is not related to redshift. In addition, galaxies are much more complex than single stellar populations, so that the "age" of a galaxy unfortunately is only an average over the different stellar populations and does not characterize its complete evolutionary stage.

Considering now $\widetilde{X}=(1+z)$, we take from \citet{Saglia2010} that $p\simeq -0.5$ and $s\simeq 0.4$. Note that these values are obtained at fixed mass, which probably biases significantly the derived evolutions of $r_e$ and $\sigma$. Anyhow if a tight fundamental plane correlation exists with the parameters previously used (those given in \citet{Saglia2010}, close to ours, do not change the main result here), then equation~(\ref{eq:condition1obs}) yields :

\begin{equation}
\left\{
\begin{array}{rcl}
     r_e    & \propto & (1+z)^{-0.5} \\
  \sigma & \propto &  (1+z)^{0.4} \\
           L &\propto &   (1+z)^{0.25} \\
\end{array}
\right.
\label{eq:redshift}
\end{equation} 

Our luminosity evolution is clearly weaker than the one estimated in  \citet{Saglia2010}: $L\propto (1+z)^1$. This discrepancy could probably be explained by the various hypotheses they have to make to try disentangling all effects in such a multivariate and evolutionary context. In particular, the evolution of $r_e$ and $\sigma$ are computed at fixed mass. possibly introducing interdependencies of variables through the $M/L$ ratio. More importantly, the cosmological clock $(1+z)$ is not a good evolutionary clock for a mixture of different populations of galaxies since they do not evolve at the same time, at the same space and along the same paths. It is well known that the tilt of the fundamental plane depends on the sample \citep{DOnofrio2008} or on the group considered \citep{Fraix2010}.

Anyhow, diversification cannot be summarized with only one simple property (like redshift or mass) because galaxies are too complex objects and do not evolve linearly in a unique way. In some diagrams, that is for some set of variables, a particular property could crudely depict the general trend of diversification. In the case of $r_e$, $\sigma$ and \mue, and to a first approximation, mass could well represent a satisfactory driving parameter for the fundamental plane correlation, but it is certainly not unique as shown in Sect.~\ref{novirial}. Since it is only approximate, some dispersion is expected.

A lot of observables evolve with diversification, at least statistically, so that we should not be surprised by the many scaling relations found for galaxies and the difficulty to pinpoint the driving parameters and mechanisms. We also better understand why several characteristic parameters (mass, luminosity, metallicity...)  and also the samples themselves have been found to influence the shape of the fundamental plane without providing a clearer picture of its origin. 

This might also explain some of the observed dispersion is most scatter plots. For instance, it has been found that the dispersion of the fundamental plane strongly depends on the evolutionary group \citep{Fraix2010}, the correlation equation~(\ref{eq:fundplane}) even not holding in the least diversified groups. Also relations several parameters like in equation~(\ref{eq:relcomb}) are most often wrong because the evolutionary clock $T$ makes most variables to be non independent. Hence, dispersion may be explained by the statistical (non-causal) nature of the correlation and the heterogeneity of the samples as far as diversification is concerned.

Indeed, the evolutionary clock, i.e. the factor $\widetilde{X}=T$, can be hidden, not understandable analytically and not directly observable. It is more directly related to an evolutionary classification, and is a well-known problem of comparative methods in phylogeny \citep[e.g.][]{Felsenstein1985}.

\section{Conclusion}

The fundamental plane correlation and similarly scaling relations for galaxies can be formalized as confounding correlations. The confounding factor $\widetilde{X}$ is very probably related to the level of diversification and may be identified in some cases with some variables that trace the global evolution of galaxies. The dependence of the variables involved in the observed correlations on $\widetilde{X}$ has been assumed here to be power law functions for sake of simplicity but they could be more complicated without changing the result of the present paper. In particular, these functions can be multivariate, with several confounding factors.

The physics thus should not be invoked to explain causally each of the observed correlations, but rather to explain the confounding or evolutionary nature of these correlations. Since the galaxy assembly history and transformation processes are complex, it is quite improbable that a simple physical theory can yield $\widetilde{X}$ and the dependence of observables on this parameter. Indeed, these functions more probably come from the statistics in the course of galaxy diversification. In addition, $\widetilde{X}$ is probably hidden, not understandable analytically and not directly observable. It might also be different depending on the set of variables and the group of galaxies considered.

Gaining insights on the confounding parameter $\widetilde{X}$ requires several complementary statistical approaches. A first one is to combine several scaling relations and several galaxy samples in order to identify some common variables that could play the role of the confounding parameter. A second approach is to use numerical simulations to produce synthetic populations of galaxies with as many assembly configurations as possible \citep[like in e.g.][]{Robertson2006, Hopkins2008, Hopkins2009}. The great advantage here is to play with unobservable parameters. Finally, the third approach is to group galaxies according to their assembly history. Since the confounding parameter $\widetilde{X}$ is probably mainly linked to the level of diversification, the scaling relations necessarily depend on the evolutionary groups. Indeed, the very nature of $\widetilde{X}$ and the functions characterizing the dependence of the variables on $\widetilde{X}$ are expected to depend on the assembly history of galaxies.

Combining evolutionary classifications, numerical simulations, observed scaling relations and recognizing the latter as evolutionary correlations, will lead us toward a much better understanding of the history of galaxy formation. Interestingly, all astrophysical objects evolve, and evolutionary correlations could probably also explain scaling relations for other stellar systems such as globular clusters \citep[e.g.][]{Misgeld2001,FDC09}.

\section*{Acknowledgments}
I would like to thank my colleagues Asis Chattopadhyay, Tanuka Chattopadhyay, Emmanuel Davoust and Marc Thuillard for numerous and enlightning discussions. 


\begin{thebibliography}{17}
\expandafter\ifx\csname natexlab\endcsname\relax\def\natexlab#1{#1}\fi

\bibitem[{{Djorgovski} \& {Davis}(1987)}]{Djorgovski1987}
{Djorgovski} S., {Davis} M., 1987, \apj, 313, 59

\bibitem[{{D'Onofrio} {et~al.}(2008){D'Onofrio}, {Fasano}, {Varela}, {Bettoni},
  {Moles}, {Kj{\ae}rgaard}, {Pignatelli}, {Poggianti}, {Dressler}, {Cava},
  {Fritz}, {Couch}, \& {Omizzolo}}]{DOnofrio2008}
{D'Onofrio} M., {Fasano} G., {Varela} J., {Bettoni} D., {Moles} M.,
  {Kj{\ae}rgaard} P., {Pignatelli} E., {Poggianti} B., {Dressler} A., {Cava}
  A., {Fritz} J., {Couch} W.~J., {Omizzolo} A., 2008, \apj, 685, 875, 0804.1892

\bibitem[{{Dressler} {et~al.}(1987){Dressler}, {Lynden-Bell}, {Burstein},
  {Davies}, {Faber}, {Terlevich}, \& {Wegner}}]{Dressler1987}
{Dressler} A., {Lynden-Bell} D., {Burstein} D., {Davies} R.~L., {Faber} S.~M.,
  {Terlevich} R., {Wegner} G., 1987, \apj, 313, 42

\bibitem[{Felsenstein(1985)}]{Felsenstein1985}
Felsenstein J., 1985, The American Naturalist, 125, pp. 1

\bibitem[{{Fraix-Burnet} {et~al.}(2009){Fraix-Burnet}, {D}avoust, \&
  {C}harbonnel}]{FDC09}
{Fraix-Burnet} D., {D}avoust E., {C}harbonnel C., 2009, MNRAS, 398, 1706,
  arXiv:0906.3458

\bibitem[{{Fraix-Burnet} {et~al.}(2010){Fraix-Burnet}, {Dugu{\'e}},
  {Chattopadhyay}, {Chattopadhyay}, \& {Davoust}}]{Fraix2010}
{Fraix-Burnet} D., {Dugu{\'e}} M., {Chattopadhyay} T., {Chattopadhyay} A.~K.,
  {Davoust} E., 2010, MNRAS, 407, 2207, arXiv:1005.5645

\bibitem[{{Gargiulo} {et~al.}(2009){Gargiulo}, {Haines}, {Merluzzi}, {Smith},
  {Barbera}, {Busarello}, {Lucey}, {Mercurio}, \& {Capaccioli}}]{Gargiulo2009}
{Gargiulo} A., {Haines} C.~P., {Merluzzi} P., {Smith} R.~J., {Barbera} F.~L.,
  {Busarello} G., {Lucey} J.~R., {Mercurio} A., {Capaccioli} M., 2009, \mnras,
  397, 75, 0902.4383

\bibitem[{Graves {et~al.}(2009)Graves, Faber, \& Schiavon}]{Graves2009}
Graves G.~J., Faber S.~M., Schiavon R.~P., 2009, The Astrophysical Journal,
  698, 1590, arXiv:0903.3603

\bibitem[{G{\"u}ltekin {et~al.}(2009)G{\"u}ltekin, Richstone, Gebhardt, Lauer,
  Tremaine, Aller, Bender, Dressler, Faber, Filippenko, Green, Ho, Kormendy,
  Magorrian, Pinkney, \& Siopis}]{Gultekin2009}
G{\"u}ltekin K., Richstone D.~O., Gebhardt K., Lauer T.~R., Tremaine S., Aller
  M.~C., Bender R., Dressler A., Faber S.~M., Filippenko A.~V., Green R., Ho
  L.~C., Kormendy J., Magorrian J., Pinkney J., Siopis C., 2009, The
  Astrophysical Journal, 698, 198, arXiv:0903.4897

\bibitem[{{Hopkins} {et~al.}(2008){Hopkins}, {Cox}, \&
  {Hernquist}}]{Hopkins2008}
{Hopkins} P.~F., {Cox} T.~J., {Hernquist} L., 2008, \apj, 689, 17,
  arXiv:0806.3974

\bibitem[{{Hopkins} {et~al.}(2009){Hopkins}, {Hernquist}, {Cox}, {Keres}, \&
  {Wuyts}}]{Hopkins2009}
{Hopkins} P.~F., {Hernquist} L., {Cox} T.~J., {Keres} D., {Wuyts} S., 2009,
  \apj, 691, 1424, arXiv:0807.2868

\bibitem[{{Marconi} \& {Hunt}(2003)}]{Marconi2003}
{Marconi} A., {Hunt} L.~K., 2003, \apjl, 589, L21, arXiv:astro-ph/0304274

\bibitem[{{Misgeld} \& {Hilker}(2011)}]{Misgeld2001}
{Misgeld} I., {Hilker} M., 2011, ArXiv e-prints, arXiv:1103.1628

\bibitem[{{Nieto} {et~al.}(1990){Nieto}, {Davoust}, {Bender}, \&
  {Prugniel}}]{Nieto1990}
{Nieto} J.-L., {Davoust} E., {Bender} R., {Prugniel} P., 1990, \aap, 230, L17

\bibitem[{{Nigoche-Netro} {et~al.}(2009){Nigoche-Netro}, {Ruelas-Mayorga}, \&
  {Franco-Balderas}}]{Nigoche-Netro2009}
{Nigoche-Netro} A., {Ruelas-Mayorga} A., {Franco-Balderas} A., 2009, \mnras,
  392, 1060, 0805.1142

\bibitem[{{Robertson} {et~al.}(2006){Robertson}, {Cox}, {Hernquist}, {Franx},
  {Hopkins}, {Martini}, \& {Springel}}]{Robertson2006}
{Robertson} B., {Cox} T.~J., {Hernquist} L., {Franx} M., {Hopkins} P.~F.,
  {Martini} P., {Springel} V., 2006, \apj, 641, 21, arXiv:astro-ph/0511053

\bibitem[{{Saglia} {et~al.}(2010){Saglia}, {S{\'a}nchez-Bl{\'a}zquez},
  {Bender}, {Simard}, {Desai}, {Arag{\'o}n-Salamanca}, {Milvang-Jensen},
  {Halliday}, {Jablonka}, {Noll}, {Poggianti}, {Clowe}, {De Lucia},
  {Pell{\'o}}, {Rudnick}, {Valentinuzzi}, {White}, \& {Zaritsky}}]{Saglia2010}
{Saglia} R.~P., {S{\'a}nchez-Bl{\'a}zquez} P., {Bender} R., {Simard} L.,
  {Desai} V., {Arag{\'o}n-Salamanca} A., {Milvang-Jensen} B., {Halliday} C.,
  {Jablonka} P., {Noll} S., {Poggianti} B., {Clowe} D.~I., {De Lucia} G.,
  {Pell{\'o}} R., {Rudnick} G., {Valentinuzzi} T., {White} S.~D.~M., {Zaritsky}
  D., 2010, \aap, 524, A6+, arXiv:1009.0645

\end{thebibliography}

\appendix

\bsp

\label{lastpage}

\end{document}